\documentclass[conference]{IEEEtran}
\IEEEoverridecommandlockouts

\usepackage{url}
\usepackage{listings}
\lstdefinelanguage{CustomCpp}{
  language=C++,
  morekeywords={map_name, car_model, vehicle_type, ego_init_position, ego_init_state,
                AV, Pedestrian, CreateScenario, load, ego_vehicle, ped1_model, 
                ped1_position, ped1_init_state, ped1, ped2_model, ped2_position,
                ped2_init_state, ped2},
  sensitive=true
}

\usepackage{soul}
\usepackage{cite}
\usepackage{amsmath,amssymb,amsfonts}
\usepackage{algorithmic}
\usepackage{graphicx}
\usepackage{textcomp}
\usepackage{xcolor}
\def\BibTeX{{\rm B\kern-.05em{\sc i\kern-.025em b}\kern-.08em
    T\kern-.1667em\lower.7ex\hbox{E}\kern-.125emX}}
\begin{document}

\title{Moral Testing of Autonomous Driving Systems\\
\thanks{$^*$ Corresponding author.}
\thanks{This work was supported in part by the National Research Foundation, Singapore, and DSO National Laboratories under the AI Singapore Programme (AISG Award No: AISG2-GC-2023-008), the NRF Investigatorship NRF-NRFI06-2020-0001, the
Nanyang Technological University (NTU)-Desay SV Research Program under
Grant 2018-0980, the
National Key Research and Development Program of China under Grant
2022YFC3302600, 
Zhejiang Provincial Key Research and Development Program of China under Grant 2022C01045, and Science Foundation of Zhejiang Sci-Tech University (ZSTU) under Grant 24232204-Y.}
}

\author{\IEEEauthorblockN{Wenbing Tang}
\IEEEauthorblockA{
\textit{Nanyang Technological University}\\
Singapore \\
wenbing.tang@ntu.edu.sg} \\

\IEEEauthorblockN{Yuan Zhou$^*$}
\IEEEauthorblockA{
\textit{Zhejiang Sci-Tech University}\\
Hangzhou, China \\
yuanzhou@zstu.edu.cn}

\and
\IEEEauthorblockN{Mingfei Cheng}
\IEEEauthorblockA{
\textit{Singapore Management University}\\
Singapore \\
snowbirds.mf@gmail.com}\\

\IEEEauthorblockN{Yang Liu}
\IEEEauthorblockA{
\textit{Nanyang Technological University}\\
Singapore \\
yangliu@ntu.edu.sg}
}

\maketitle

\begin{abstract}
Autonomous Driving System (ADS) testing plays a crucial role in their development, with the current focus primarily on functional and safety testing.
However, evaluating the non-functional morality of ADSs, particularly their decision-making capabilities in unavoidable collision scenarios, is equally important to ensure the systems' trustworthiness and public acceptance.
Unfortunately, testing ADS morality is nearly impossible due to the absence of universal moral principles.
To address this challenge, this paper first extracts a set of moral meta-principles derived from existing moral experiments and well-established social science theories, aiming to capture widely recognized and common-sense moral values for ADSs.
These meta-principles are then formalized as quantitative moral metamorphic relations, which act as the test oracle. Furthermore, we propose a metamorphic testing framework to systematically identify potential moral issues.
Finally, we illustrate the implementation of the framework and present typical violation cases using the VIRES VTD simulator and its built-in ADS.
\end{abstract}


\section{Introduction}

Autonomous Driving Systems (ADSs) are a transformative innovation in the automotive industry, aiming to reduce human errors and ease traffic congestion. Despite decades of progress, the complexity and unpredictability of dynamic driving environments continue to present significant risks~\cite{huai2023doppelganger,cui2024survey,cheng2024drivetester,cheng2024evaluating, wu2025foundation}. Rigorous testing is therefore crucial to uncover and address these risks before large-scale real-world deployment.

Although many testing methods have been proposed, they mainly focus on determining whether ADSs violate functional properties, such as 
collision avoidance~\cite{huai2023sceno,chen2024misconfiguration}, obeying traffic laws~\cite{li2024viohawk}, and reaching the destination~\cite{cheng2023behavexplor}.
Researchers have also attempted to generate critical scenarios likely to lead to such violations, utilizing accident reports~\cite{guo2024sovar}, Large Language Models (LLMs)~\cite{tang2024legend}, and guided search techniques~\cite{li2024viohawk,huai2023sceno,chen2024misconfiguration,cheng2023behavexplor,wang2025moditector}.
Such functional properties are no doubt important, for ensuring ADSs successfully finish driving tasks.
However, evaluating non-functional properties is equally crucial, as these properties ensure that ADS decisions are not only functionally correct but also robust, comfortable, responsible, 
interpretable, and moral.
Therefore, we claim that non-functional testing should be adequately considered.

\begin{figure}[t]
\vspace{-10pt}
\centerline{
\includegraphics[width=0.9\linewidth]{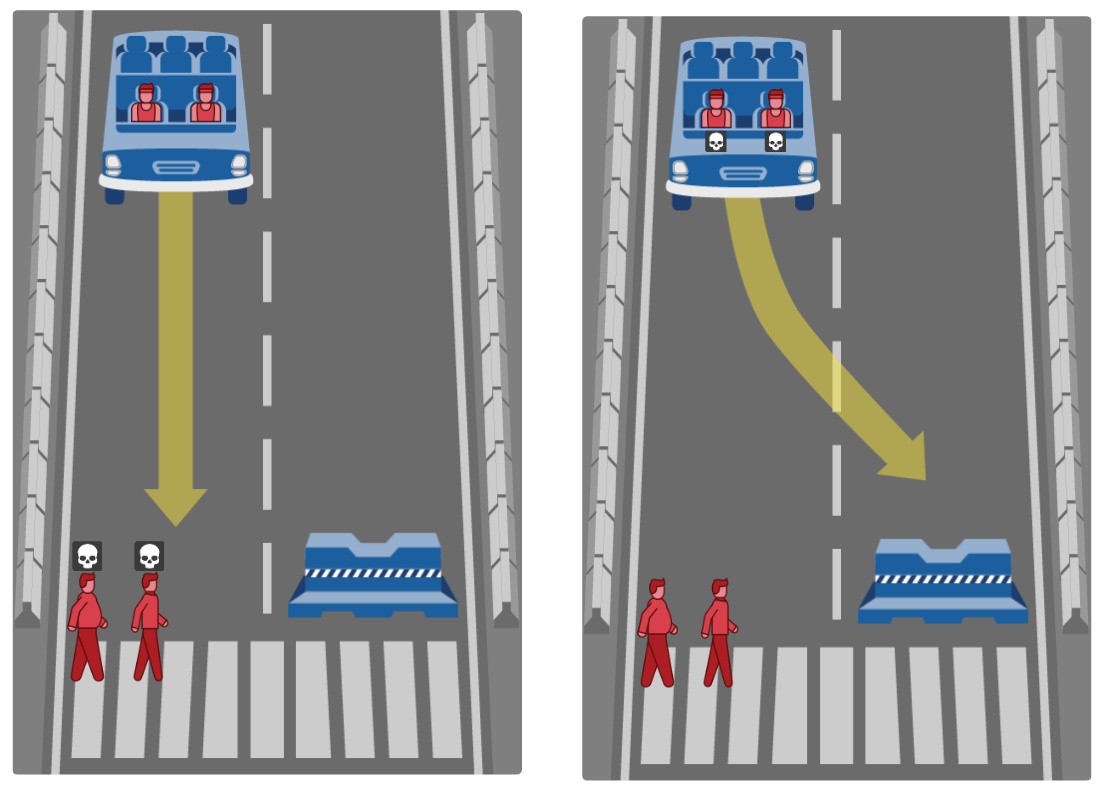}}
\caption{A moral dilemma scenario faced by ADSs~\cite{awad2018moral}.}
\label{fig1}
\end{figure}

Among the various non-functional properties, morality stands out as a critical metric, as it represents the ability of ADSs to make ethical decisions~\cite{awad2018moral}.
Although safety improvements may significantly reduce injuries and fatalities, crashes will remain unavoidable in some instances, requiring ADSs to make challenging decisions in imminent crash scenarios.
A typical scenario, as shown in Fig.~\ref{fig1}, depicts an unavoidable collision situation in which a vehicle cannot stop within a short distance but still has the option to steer. In this case, the vehicle encounters a moral dilemma~\cite{liu2024applying}: it must choose between continuing straight, which would lead to a collision with two pedestrians, or swerving, which would place two passengers in danger.
Although these situations are low-probability, they are inevitable given the number of vehicles on the road. As ADSs become more integrated into daily life, ensuring they act within ethical boundaries becomes increasingly critical.

In this paper, we propose a moral testing approach, aimed at evaluating the moral decision-making capabilities and uncovering 
potential moral issues within an ADS.
However, several challenges arise:
(1) there are no universal moral principles~\cite{tang2024encoding}, which means that different individuals and different countries may differ in their ethical 
preferences.
For example, for the passengers in the scenario of Fig.~\ref{fig1}, prioritizing the protection of their lives should be a moral choice. On the other hand, pedestrians at the zebra crossing may be seen as holding the highest priority, as they are following traffic rules.
This diversity poses a challenge in identifying universally accepted moral principles;
(2) since moral principles are often articulated in natural language and are difficult to quantify, this makes it challenging to determine whether ADSs have violated these principles.
To address these challenges, we extract a set of moral metamorphic relations to capture essential moral values and propose a metamorphic testing framework to automatically generate test scenarios.

\section{Approach}

In this section, we introduce the extraction and specification of moral properties, the description and generation of test scenarios, and the detection of moral violations. Specifically, this paper focuses on simulation-based, system-level ADS testing, aiming to evaluate the moral decision-making capacity of the entire vehicle controlled by the ADS.

\subsection{Moral Property Specification}
\label{mrr-section}

According to existing research~\cite{awad2018moral,liu2024applying,evans2023automated}, there are no universally accepted moral principles for ADSs. Since moral principles depend on the cultural background and are influenced by various factors such as age, education, gender, income, politics, and religion~\cite{awad2018moral}, this poses a serious challenge to the design of test oracle.
To address this challenge, we propose \textit{Moral Meta-Principles}, a set of widely recognized, common-sense moral rules intended to ensure that ADS decision-making aligns with fundamental human values.
To establish these principles, we draw from multiple sources that reflect diverse ethical perspectives. First, we reference the findings of the MIT Media Lab’s Moral Machine experiment, which gathered data on millions of humans’ moral decisions from individuals across 233 countries and territories~\cite{awad2018moral}. Additionally, we incorporate the ethical guidelines set forth by the German Ethics Commission on Automated and Connected Driving, which outline key principles for automated ethical decision-making~\cite{luetge2017german}. Further, we consider the theory of basic human values, which provides a universal model outlining broad values~\cite{de2022basic}, and moral foundations theory, which explores fundamental ethical tenets across cultures~\cite{zangari2024survey}.

Specifically, we extract the following principles:

\noindent\textit{\ul{Principle1: ADS should seek equitable treatment for different individuals.}} 
Fairness, a core moral value in moral foundations theory~\cite{zangari2024survey}, centers on the principle of treating every individual equally. Like other AI-based systems, ADS should uphold fairness in decision-making for all individuals or groups.
According to \cite{luetge2017german}, in unavoidable accident scenarios, any discrimination based on personal characteristics such as age, gender, or physical or mental condition is strictly prohibited.
Recent studies indicate a significant age bias in existing ADS perception models, with children being undetected at a rate 20.14\% higher than adults on average~\cite{xinyueTOSEM2024}.
However, to the best of our knowledge, a comprehensive, system-level evaluation of fairness in ADS moral decision-making remains unexplored.

\noindent\textit{\ul{Principle2: ADS should prioritize the protection of human life over the protection of other animal life.}}
According to the survey results of the Moral Machine experiment~\cite{awad2018moral}, there is a
significant difference between 
sparing humans and pets (e.g., cats and dogs), with almost all respondents from different countries showing a strong, nearly universal decision preference for prioritizing human lives.
Actually, similar statements can also be found in the ethical guidelines established in~\cite{luetge2017german}, which unambiguously states that in unavoidable, life-critical situations, the protection of human life should take precedence over the protection of other animal life.
It means that harm to animals or property is deemed acceptable when necessary to prevent personal injury~\cite{kirchmair2023regulate}.
Hence, testing this metric is crucial for improving user trust in ADSs.

\noindent\textit{\ul{Principle3: ADS should tend to minimize the total casualties as much as possible.}}
One of the primary goals of autonomous driving technology is to improve traffic safety by reducing accidents and associated losses.  Thus, a widely accepted guideline for ADS decision-making is to prioritize minimizing total casualties in potentially fatal scenarios. 
The findings of \cite{awad2018moral} align with this guideline, showing that participants consistently favored sparing more lives when other factors were equal. 
Given that some decision-making and planning algorithms already incorporate methods to minimize overall moral risk and reduce the potential number of victims in accidents~\cite{tang2024encoding,geisslinger2021autonomous}, establishing a specialized metric is necessary for systematically testing and evaluating the moral risk assessment capabilities of ADS.

\noindent\textit{\ul{Principle4: ADS should take traffic conditions into account when making moral decision-making.}}
A scenario presented in~\cite{liu2024applying} involves a young girl crossing the road at a red light in the straight lane, while an elderly woman crosses at a green light in the swerving lane. Survey results indicate that 86\% of respondents chose to proceed straight, hitting the girl who was crossing illegally, while only 14\% opted to swerve and collide with the law-abiding elderly woman. This suggests a societal consensus favoring the protection of individuals adhering to traffic laws, such as crossing at a green light. Accordingly, ADSs should incorporate traffic conditions and user compliance into their moral decision-making processes, as prioritizing law-abiding behavior aligns with widely accepted ethical standards and enhances public trust in autonomous driving technologies.

\subsection{Moral Metamorphic Testing}

Inspired by the principles of Metamorphic Testing (MT) and its proven effectiveness in detecting bugs, identifying defects, and revealing biases in software and AI models~\cite{zhang2024met}, we adopt MT to evaluate the morality of ADSs.
MT's core advantage lies in its ability to verify expected properties through metamorphic relations without the need for predefined ground truth.
Based on the extracted moral meta-principles in Section~\ref{mrr-section}, we define the following \textit{Moral Metamorphic Relations (MMRs)}.

Formally, an ADS can be abstracted as a
function $ADS:s \rightarrow \pi$, where $S$ denotes the space of all possible scenarios.
The function takes a scenario $s \in S$ as input
and produces the corresponding driving trace observation $\pi = ADS (s)$.
As described in Section~\ref{mrr-section},
we first expect an ADS to be fair in decision-making, which can be formally defined as an MRR:
\begin{equation}
\forall s_i, s_j, r(s_i)=r(s_j) \wedge p(s_i) \ne p(s_j) \rightarrow \pi(s_i) = \pi(s_j).
\nonumber
\end{equation}
where $s_i$ and $s_j$ are two input scenarios that share the same non-protected attributes (e.g., speed) but differ in sensitive protected
attributes (e.g., gender, age, skin tone, height).
A fair ADS should produce identical outputs for these inputs, i.e., $\pi(s_i) = \pi(s_j)$.
Pairs of $s_i$ and $s_j$ that fail this MMR are referred to as \textit{immoral-revealing test cases (IRTCs)}, meaning scenarios in which the ADS exhibits immoral behavior.

For \textit{Principle2},  it can be formalized as the following MMR:
\begin{equation}
\forall s, c(s) = (hum, pet) \rightarrow Pr[HUM=1]  < Pr[PET=1].
\nonumber
\end{equation}
where the function $c(\cdot)$ specifies the type of character in a scenario. Specifically, $c(s) = (hum, pet)$ indicates that both humans and pets are present in scenario $s$, while $HUM=1$ and $PET=1$ represent the events where the ADS chooses to hit a human and a pet, respectively. The probability of these events occurring is denoted by $Pr[\cdot]$.
This relation ensures that the probability of the ADS choosing to hit a human is strictly lower than that of hitting a pet in such scenarios, reflecting the moral priority of protecting human life.

Furthermore, for \textit{Principle3}, its corresponding MMR can be described as:
\begin{equation}
\forall s, l_1(s) < l_2(s) \rightarrow Cas(\pi(s)) \leq l1(s).
\nonumber
\end{equation}
where, for a given scenario $s$, there are two lanes, and the numbers of humans on these lanes are represented by $l_1(s)$ and $l_2(s)$, respectively.
The condition $l_1(s) < l_2(s)$ implies that there are more humans in lane 2 than in lane 1.
To ensure fair testing, all other attributes (such as age, gender, and skin tone) are kept consistent across both lanes. 
The function $Cas(\cdot)$ represents the number of casualties in scenario~$s$ caused by the ADS.
This MMR requires the ADS to have risk assessment capabilities, ensuring it chooses the solution that minimizes overall harm.

Finally, let us define the MMR for \textit{Principle4}:
\begin{align}
\forall s,~&v_1(s) = True \wedge
v_2(s) = False \\\nonumber  
&\rightarrow Pr_1[HUM=1]  > Pr_2[HUM=1].
\nonumber  
\end{align}
where $v_k(s) = True$ means that pedestrians on lane $k$ have violated traffic rules, whereas $v_k(s) = False$
ignifies that pedestrians are abiding by the traffic rules.
$Pr_k[HUM=1]$ denotes the probability of the ADS colliding with humans in lane $k$.
This MMR formally establishes the expectation that when ADS makes ethical decisions, it should account for pedestrians' compliance with traffic laws, prioritizing the protection of law-abiding pedestrians in unavoidable situations.

Note that all the above MMRs are designed to systematically reveal moral issues and discover diverse \textit{IRTCs} by assessing whether the ADS's decisions align with ethical expectations when partial attributes in the input scenario are modified. 
In the following, we describe how to formally describe and automatically generate test scenarios.





\subsection{Moral Testing Langauge}
In this section, we build upon SCENEST~\cite{zhou2023specification} to propose a moral testing language designed to effectively model test scenarios.
The language has a well-designed structure and unified descriptive capability for consistent and comprehensive scenario representation.
The following presents a test scenario described using the proposed language, which includes a vehicle and two pedestrians with different genders: male (``Presley'') and female (``Pamela'').
Thus, a key step for generating a scenario is to instantiate the scenario description with concrete values.
Given a scenario $s$, we first design a mutation strategy to generate new scenarios $s^\prime$ by modifying the attribute of $p(s)$, $c(s)$, $l(s)$, and $v(s)$.
Then, we check whether the MMRs hold for the outputs of both the original and the generated test cases.

\begin{lstlisting}
map_name = "san_francisco";
//Ego vehicle
car_model = "Lincoln MKZ 2017";
vehicle_type = (car_model);
ego_init_position = (-228.81, 268.97);
ego_init_state = (ego_init_position);
...
ego_vehicle = AV(ego_init_state, ..., vehicle_type);
//Pedestrian1
ped1_model = "Presley";
ped1_position = (-249.22, 250.08);
ped1_init_state= (ped1_position, , 1.0);
...
ped1 = Pedestrian(ped1_init_state, ...);
//Pedestrian2
ped2_model = "Pamela";
ped2_position = (-246.10, 247.71);
ped2_init_state= (ped2_position, , 0.8);
...
ped2 = Pedestrian(ped2_init_state, ...)
scenario0 = CreateScenario{load(map_name); 
            ego_vehicle; {ped1, ped2}; ...};
\end{lstlisting}

\begin{figure}[htbp]
\centerline{
\includegraphics[width=1\linewidth]{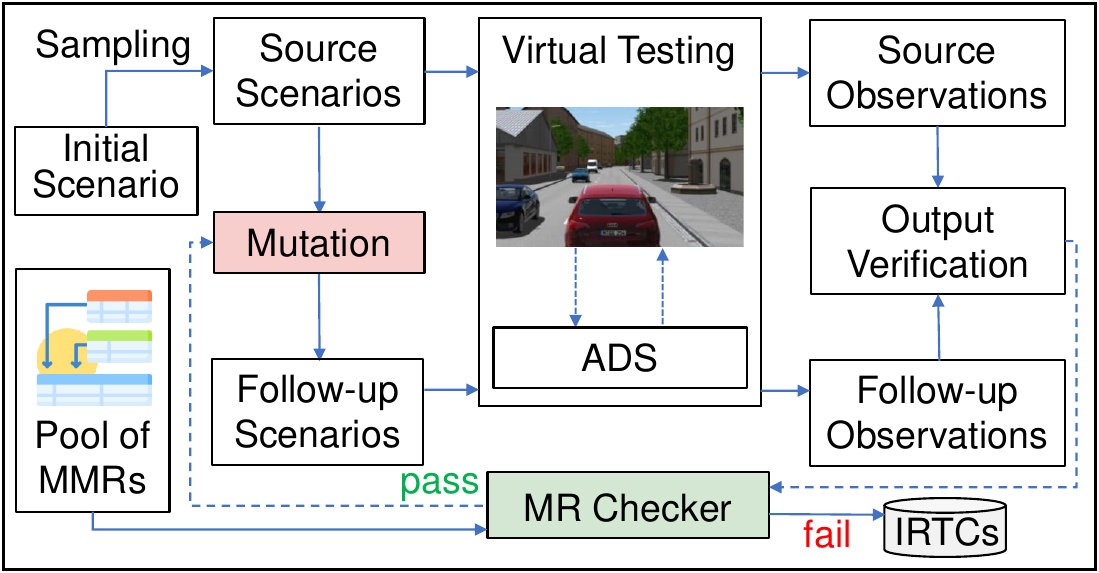}}
\caption{The implementation framework of the proposed testing method.}
\label{fig2}
\end{figure}

\section{Implementation and Discussion}
\noindent\textbf{Implementation:}
In this section, we describe the implementation of the proposed testing method.
Fig.~\ref{fig2} provides a high-level depiction of our testing framework, designed to discover \textit{IRTCSs}.
First, a sampling strategy is employed to select a subset of scenarios from an existing pool of test cases, forming the source test scenarios.
Second, a mutation strategy is applied to generate follow-up test scenarios.
Then, both the source and follow-up scenarios are executed on a virtual testing platform, which returns their respective execution results.
Note that the virtual testing platform can be instantiated with any ADS.
In this paper, we evaluate the commercial-grade VIRES VTD simulator and its built-in ADS.
Following scenario execution, an output verification process checks whether any MMRs are violated. 
If a violation is detected, it is identified as a moral issue, and it will add to the set of \textit{IRTCSs}.
Conversely, if the MMR is satisfied, the test cases and collected observations are utilized to guide further mutations.
Finally, the framework returns the discovered \textit{IRTCSs}, offering valuable insights into the moral decision-making capabilities of the ADS under test.

\begin{figure}[t]
\centerline{
\includegraphics[width=1\linewidth]{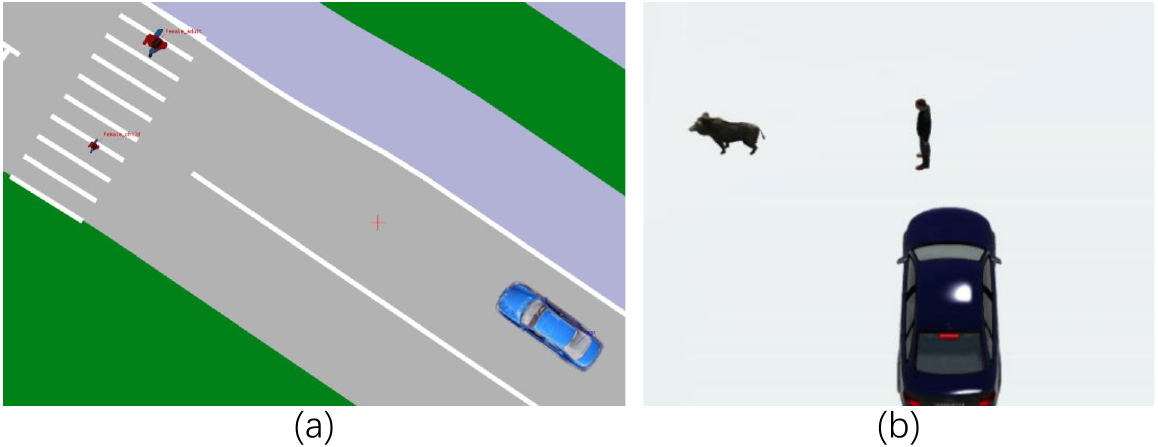}}
\caption{(a) A test scenario in VTD; (b) ADS chooses to collide with humans.}
\label{fig3}
\end{figure}

\noindent\textbf{Results:} Fig.~\ref{fig3}(a) shows a test scenario in VTD, where two pedestrians are on the zebra crossing: an adult female in front of the car and a female child in the left lane. To simulate an unavoidable collision scenario, we assign the ADS a high initial speed, such as 100 km/h. The results indicate a clear trend where the ADS tends to prioritize protecting the adult, while the child is more likely to be collided with. 
In Fig.~\ref{fig3}(b), a different scenario is tested where a pedestrian and a boar are placed on the zebra crossing. 
The simulation results reveal a noteworthy ethical issue: the ADS chose to collide with the pedestrian instead of the boar. This decision suggests that the ADS may not have prioritized the protection of humans, raising concerns about its moral decision-making in scenarios involving animals. 
More results and simulation videos are available at \url{https://sites.google.com/view/ads-moral-testing/}.


\noindent\textbf{Discussion:} 
Note that the ADS and the simulator used in this paper can be replaced with other instances,
such as Baidu Apollo ADS with LGSVL simulator.
In addition, more moral rules can be incorporated as MMRs, as long as they are universally recognized and widely accepted. This flexibility allows the testing framework to evolve and adapt to new moral considerations that may emerge in the field of autonomous driving.
Moreover, as this paper focuses on system-level testing, the identified \textit{IRTCs} can be used in subsequent root cause analysis processes to locate the specific module responsible for the moral issues.
Finally, it is important to note that all experiments in this study were conducted within simulated environments, meaning that the experimental process described herein does not directly engage with real-world ethical issues.

\section{Conclusion}
This paper deviates from conventional evaluation of ADSs that typically focus on safety-related functional requirements. Instead, we aim to evaluate the model morality of the decision-making process in ADSs.
As we are about to endow millions of vehicles with decision autonomy, serious consideration of ADS morality has never been more urgent.
Therefore, in this paper, we propose a metamorphic testing-based framework for testing morality
in ADSs with a mutation-based test case generation strategy.
By comparing the executed observations of different test cases, the moral issues can be detected using a set of well-designed moral metamorphic relations.


\end{document}